\documentclass[aps,12pt,superscriptaddress,amsfonts,amssymb,amsmath]{revtex4-2}
\pdfoutput=1

\usepackage{graphicx}
\usepackage{epsfig}
\usepackage{makeidx}
\usepackage{xcolor}
\usepackage{amsmath}
\usepackage{bm}
\usepackage{amssymb}

\begin{document}

\title{Quantum collapse, local conservation of charge, and possible experimental consequences} 

\author{F.\ Minotti \footnote{Email address: minotti@df.uba.ar}}
\affiliation{Universidad de Buenos Aires, Facultad de Ciencias Exactas y Naturales, Departamento de F\'{\i}sica, Buenos Aires, Argentina}
\affiliation{CONICET-Universidad de Buenos Aires, Instituto de F\'{\i}sica Interdisciplinaria y Aplicada (INFINA), Buenos Aires, Argentina}

\author{G.\ Modanese \footnote{Email address: giovanni.modanese@unibz.it}}
\affiliation{Free University of Bozen-Bolzano \\ Faculty of Engineering \\ I-39100 Bolzano, Italy}
\date{\today}

\linespread{0.9}

\begin{abstract}
 
We investigate the possibility that idealized quantum state-reduction processes may produce a local violation of charge conservation. If this occurs, the corresponding electromagnetic fields cannot be consistently described within Maxwell electrodynamics, and a natural alternative is provided by Aharonov-Bohm electrodynamics, which reduces to Maxwell theory when local charge conservation holds, but remains compatible with non-conserved sources. Within this framework we first analyze how state reduction may generate non-conserved local currents, including statistically compensated cases and biased tunnelling configurations with persistent average current. We then study the interaction of gauge waves with fermionic and bosonic quantum systems, the latter being described by a modified Schrödinger equation previously proposed for boson matter. As an application, we discuss the interaction of gauge waves with superconductors and show that they can effectively shield such waves. Finally, we present experimental proposals based on inverse-biased diodes and estimate the expected detector response.
 
\end{abstract}

\maketitle

\section{Introduction}

From Schr\"{o}dinger equation one can derive the expression of the locally conserved electric current, which can as well be obtained as the Noether current from the global gauge invariance of the action for the quantum system interacting with the electromagnetic (e.m.) field. It thus appears that no violation of local conservation of charge is possible. However, as shown below, those derivations are valid as long as the system obeys Schr\"{o}dinger equation at all times, and thus it is legitimate to wonder what would happen during an idealized state-reduction process, which is not necessarily described by Schr\"{o}dinger equation. We argue below that the usual interpretation of a quantum collapse process does indeed lead in general to a violation of local conservation of charge, although the conservation could still hold in a statistical sense in some cases. A direct consequence of a non-conserved current is that it is incompatible with Maxwell electrodynamics, and so the e.m.\  field is to be described by an alternative theory, like that of Aharonov and Bohm (AB) \cite{aharonov1963further}. AB electrodynamics has the desirable property of reducing to Maxwell theory when charge is strictly conserved locally, while being compatible with its non-conservation, giving rise to a rich new phenomenology when this happens \cite{hively2012toward,Modanese2017MPLB,modanese2018time,Minotti-Modanese-Symmetry2021,minotti2021quantum,minotti2022electromagnetic,hively2019classical,EPJC2023,minotti_modanese_math13050892}.
An important consequence of AB theory is the existence of certain field configurations denoted as gauge waves. These are present also in Maxwell theory, but cannot be considered as physical in its context because they have no interaction with matter when local conservation holds.
In previous works \cite{minotti2021quantum,EPJC2023} we have analyzed how these waves interact with classical systems with local non-conservation. In the present work we extend some of those results to the case of their interaction with a quantum system.

For the reader’s convenience, let us briefly recall the basic structure of Aharonov-Bohm electrodynamics. In this theory the fundamental electromagnetic variable is the four-potential $A^\mu=(\varphi/c,\mathbf{A})$, while the field tensor is defined as usual by $F^{\mu\nu}=\partial^\mu A^\nu-\partial^\nu A^\mu$. In addition to $F^{\mu\nu}$, the theory contains the scalar quantity
$$
 S=\partial_\mu A^\mu,
 $$
 which becomes dynamical and, in general, cannot be gauged away. The electromagnetic Lagrangian density is
 $$
 \mathcal{L}_{\rm AB}=-\frac{1}{4\mu_0}F_{\mu\nu}F^{\mu\nu}
 -\frac{1}{2\mu_0}(\partial_\mu A^\mu)^2.
$$
 Unlike Maxwell electrodynamics, this formulation is compatible with sources for which
$$
 I\equiv \partial_\mu j^\mu \neq 0,
$$
where $I$ measures the local failure of charge conservation and acts as the source of the scalar mode $S$. When $I=0$, the theory reduces to the Maxwell case for localized sources; when $I\neq0$, additional phenomena become possible, including gauge-wave components that remain physically ineffective for ordinary conserved probes, but may interact with systems in which local conservation does not hold. In this framework, the extra power and force densities acting on an anomalous probe $p$ are proportional to $-I_p\varphi$ and $-I_p\mathbf{A}$, respectively.

The paper is organized as follows. In Section II we discuss how an idealized state-reduction process may lead to a local failure of charge conservation, first at the level of single events and then in statistical or time-averaged regimes, including a biased tunnelling configuration with a persistent current. In Section III we study the interaction of a gauge wave with a fermionic quantum system, while in Section IV we extend the analysis to bosonic systems described by the modified Schrödinger equation proposed in Ref. \cite{minotti_modanese_IJMPA2026}. In Section V we apply the latter framework to superconductors and analyze the shielding of gauge waves. Section VI presents experimental proposals based on inverse-biased diodes and discusses the expected detector response. Finally, Section VII summarizes the main conclusions and outlines possible directions for future work.

\section{Quantum collapse and AB electrodynamics}
\label{QC_ABE}

We consider a general system with e.m.\
interactions whose e.m.\ fields are given in terms of the four potential $
A_{\mu }=\left( \frac{1}{c}\varphi ,-\mathbf{A}\right) $. Greek indices
denote the four coordinates, with $x_{0}=ct$, while Latin indices indicate the three spatial coordinates.

The system, matter plus e.m.\ fields, is described by an action of the generic form
\begin{equation*}
S=\int_{R}\mathcal{L}\left( \phi ,\partial _{\mu }\phi \right) d^{4}x
\end{equation*}
in which $\phi $ represents all matter and e.m.\ fields, and $R$ is a space-time region bounded by two space-like hypersurfaces \cite{ryder1994}.

The equations of motion and the conserved magnitudes are determined from this
action in the usual manner using Hamilton principle and Noether theorem. We will be concerned
only with conserved magnitudes arising from internal symmetries, so that space-time related symmetries will not be considered in the following.

Variation of the fields results in the variation of the action given by
\begin{eqnarray}
\delta S &=&\int_{R}\left[ \frac{\partial \mathcal{L}}{\partial \phi }\delta
\phi +\frac{\partial \mathcal{L}}{\partial \left( \partial _{\mu }\phi
\right) }\delta \left( \partial _{\mu }\phi \right) \right] d^{4}x  \notag \\
&=&\int_{R}\left[ \frac{\partial \mathcal{L}}{\partial \phi }-\partial _{\mu
}\frac{\partial \mathcal{L}}{\partial \left( \partial _{\mu }\phi \right) }
\right] \delta \phi d^{4}x+\int_{\partial R}\frac{\partial \mathcal{L}}{
\partial \left( \partial _{\mu }\phi \right) }\delta \phi d\sigma _{\mu },
\label{deltaS}
\end{eqnarray}
where $\partial R$ corresponds to the boundary of $R$.

The application of Hamilton principle thus results in the equations of motion for the fields
\begin{equation}
\frac{\partial \mathcal{L}}{\partial \phi }-\partial _{\mu }\frac{\partial 
\mathcal{L}}{\partial \left( \partial _{\mu }\phi \right) }=0.
\label{eqsmot}
\end{equation}

On the other hand, Noether theorem applies in the case in which a given variation of the fields, of the generic form $\delta \phi =\varepsilon \Phi $, with $\varepsilon $ an infinitesimal parameter, leaves the action invariant: $\delta S=0$. Further assuming that the given variation is that of fields that satisfy the equations of motion (on-shell condition), the action invariance means that, for a generic $\partial R$, 
\begin{equation*}
\int_{\partial R}\frac{\partial \mathcal{L}}{\partial \left( \partial _{\mu
}\phi \right) }\Phi d\sigma _{\mu }=0,
\end{equation*}
or, equivalently, $\partial _{\mu }J^{\mu }=0$, with
\begin{equation}
J^{\mu }=\frac{\partial \mathcal{L}}{\partial \left( \partial _{\mu }\phi
\right) }\Phi .  \label{Jmu}
\end{equation}

Let us consider as a concrete example a quantum system described by Schr\"{o}dinger equation, whose  matter Lagrangian with no interactions can be written as \cite{padmanabhan2018}: 
\begin{equation}
\mathcal{L}_{M}=\frac{i\hbar c}{2}\left( \psi ^{\ast }\partial _{0}\psi
-\psi \partial _{0}\psi ^{\ast }\right) -\frac{\hbar ^{2}}{2m}\nabla \psi
\cdot \nabla \psi ^{\ast }.  \label{LMSchroedinger}
\end{equation}

In \cite{minotti_modanese_IJMPA2026} we explored the possibility of a physically motivated principle for the coupling of matter to e.m.\ fields, different from the minimal coupling principle, with the conclusion that for fermions both principles are equivalent, but differ for bosons. In this section and in Section \ref{AB_QS} we consider a fermion system and so refer to the more familiar minimal coupling. The coupling to a boson system is considered in Sections \ref{GW_Boson} and \ref{GW_SC}. 

The minimal coupling principle tells us that matter of electric charge $q$ interacting with the e.m.\ field is described by the matter Lagrangian (\ref{LMSchroedinger}) in which partial derivatives are replaced by covariant ones, expressed in terms of the e.m.\ four-potential $A_{\mu}=(\varphi/c,-\mathbf{A})$  \cite{huang2010quantum},
\begin{equation*}
\partial _{\mu }\rightarrow \partial _{\mu }+i\frac{q}{\hbar }A_{\mu },
\end{equation*}
that is
\begin{eqnarray*}
\mathcal{L}_{M\_int} &=&\frac{i\hbar c}{2}\left[ \psi ^{\ast }\left(
\partial _{0}+i\frac{q}{\hbar c}\varphi \right) \psi -\psi \left( \partial
_{0}-i\frac{q}{\hbar c}\varphi \right) \psi ^{\ast }\right] \\
&&-\frac{\hbar ^{2}}{2m}\left( \nabla -i\frac{q}{\hbar }\mathbf{A}\right)
\psi \cdot \left( \nabla +i\frac{q}{\hbar }\mathbf{A}\right) \psi ^{\ast }.
\end{eqnarray*}

The full Lagrangian requires the addition of the e.m.\ Lagrangian $\mathcal{L}_{em}$, which depends only on $A_{\mu }$. It is thus readily checked that variation of the matter field in the corresponding full action 
\begin{equation}
S=\int \left( \mathcal{L}_{M\_int}+\mathcal{L}_{em}\right) d^{4}x
\label{Sfull}
\end{equation}
gives the Schr\"{o}dinger equation 
\begin{equation}
i\hbar \frac{\partial \psi }{\partial t}=H\psi ,  \label{Scheq}
\end{equation}
with
\begin{equation}
H=\frac{1}{2m}\left( -i\hbar \nabla -q\mathbf{A}\right) ^{2}\ +q\varphi
 .  \label{Hamiltonian}
\end{equation}

We further note that the action (\ref{Sfull}), with  $\mathcal{L}_{em}$ corresponding to either Maxwell, or AB electrodynamics, is invariant for a global gauge transformation of the fields given by 
\begin{subequations}
\label{global}
\begin{eqnarray}
\delta \psi &=&-i\frac{q}{\hbar }\Lambda \psi , \\
\delta \psi ^{\ast } &=&i\frac{q}{\hbar }\Lambda \psi ^{\ast }, \\
\delta A^{\mu } &=& \partial^{\mu}\Lambda=0,
\end{eqnarray} 
\end{subequations}
with $\Lambda $ an infinitesimal constant.

The corresponding conserved Noether current (\ref{Jmu}) is thus
\begin{subequations}
\label{conservedJ}
\begin{eqnarray}
J^{0} &=&qc\psi ^{\ast }\psi , \\
J^{j} &=&\frac{iq\hbar }{2m}\left( \psi \partial _{j}\psi ^{\ast }-\psi
^{\ast }\partial _{j}\psi \right) -\frac{q^{2}}{m}\psi ^{\ast }\psi A_{j},
\end{eqnarray}
\end{subequations}
a well-known result.

More explicitly, for the particular global gauge transformation (\ref{global}) the resulting condition $\delta S=0$ gives 
\begin{equation}
\int_{R} \left\{ \partial _{\mu }J^{\mu }+\frac{iq}{\hbar }\left[ \psi ^{\ast
}\left( i\hbar \frac{\partial \psi }{\partial t}-H\psi \right) -\psi \left(
i\hbar \frac{\partial \psi }{\partial t}-H\psi \right) ^{\ast }\right]
\right\} d^{4}x=0,  \label{dSzero}
\end{equation}
from which $\partial _{\mu }J^{\mu }=0$ follows if $\psi $ satisfies its equation (\ref{Scheq}), the usual ``on-shell" form of Noether theorem. 

The question is now: what would happen during an idealized state-reduction process?

We can describe such a process occurring at $t=t_{0}$ by writing the wave function as
\begin{equation}
\psi =\psi _{+}\Theta \left( t-t_{0}\right) +\psi _{-}\Theta \left(
t_{0}-t\right) ,  \label{psicollapse}
\end{equation}
where $\Theta $ is the unit step function, and $\psi _{+}$ satisfies Schr\"{o}dinger equation for $t>t_{0}$, while $\psi _{-}$ does it for $t<t_{0}$.

If we replace expression (\ref{psicollapse}) in (\ref{dSzero}) the result is
\begin{equation}
\partial _{\mu }J^{\mu }=q\left( \left\vert \psi _{+}\right\vert
^{2}-\left\vert \psi _{-}\right\vert ^{2}\right) \delta \left(
t-t_{0}\right) ,  \label{DmuJmu}
\end{equation}
where $\delta $ is Dirac delta.

It is interesting to note that (\ref{DmuJmu}) is a more general expression of the model in \cite{minotti2022electromagnetic,EPJC2023}, which is of the form
\begin{equation}
\partial _{\mu }J^{\mu }=i_{n}\left[ \delta \left( \mathbf{x}-\mathbf{x}_{n}-
\mathbf{d}_{n}\right) -\delta \left( \mathbf{x}-\mathbf{x}_{n}\right) \right]
,  \label{QR2022}
\end{equation}
where the elementary current $i_{n}$ is considered to instantaneously flow from the point $\mathbf{x}_{n}$ to the point $\mathbf{x}_{n}+\mathbf{d}_{n}$. Relation (\ref{DmuJmu}) reduces to (\ref{QR2022}) for a highly localized charge density in which 
\begin{eqnarray*}
i_{n} &=&q\delta \left( t-t_{0}\right) , \\
\left\vert \psi _{+}\right\vert ^{2} &=&\delta \left( \mathbf{x}-\mathbf{x}
_{n}-\mathbf{d}_{n}\right) , \\
\left\vert \psi _{-}\right\vert ^{2} &=&\delta \left( \mathbf{x}-\mathbf{x}
_{n}\right) .
\end{eqnarray*}

In this way, if the wave collapse process leads to $\partial _{\mu }J^{\mu}\neq 0$\ the e.m.\ part of the action cannot correspond to Maxwell Lagrangian
\begin{equation}
\mathcal{L}_{em}=-\frac{1}{4\mu _{0}}F^{\mu \nu }F_{\mu \nu },
\label{Lem_Maxwell}
\end{equation}
which requires charge-current sources that are locally conserved.

A possibility then is to consider Aharonov-Bohm's (AB) Lagrangian \cite{aharonov1963further} for the e.m.\ part:
\begin{equation}
\mathcal{L}_{em}^{AB}=-\frac{1}{4\mu _{0}}F^{\mu \nu }F_{\mu \nu }-\frac{1}{
2\mu _{0}}\left( \partial ^{\mu }A_{\mu }\right) ^{2},  \label{LemAB}
\end{equation}
which is compatible with non-conserved sources.

The AB Lagrangian can be recast as, leaving out a full divergence,
\begin{equation*}
\mathcal{L}_{em}^{AB}=-\frac{1}{2\mu _{0}}\left( \partial _{\mu }A_{\nu
}\right) \left( \partial ^{\mu }A^{\nu }\right) ,
\end{equation*}
a simpler and more convenient expression for calculations.

The AB Lagrangian is not invariant for local
gauge transformations, but it is invariant for the global gauge transformation (\ref{global}). This means that if it is used as the e.m.\ part of the full action, Noether theorem tells us that the current (\ref{conservedJ}) is conserved as long as the effect of wave collapse can be neglected. In such a case the resulting non-homogeneous equation for the e.m.\ field
\begin{equation*}
\partial ^{\nu }\partial _{\nu }A^{\mu }=\mu _{0}J^{\mu },
\end{equation*}
satisfies
\begin{equation*}
\partial ^{\nu }\partial _{\nu }\left( \partial _{\mu }A^{\mu }\right) =0,
\end{equation*}
whose solution for localized sources is $\partial _{\mu }A^{\mu }=0$.
Consequently, the AB Lagrangian coincides with Maxwell's and no difference is observed in the phenomenology between both approaches. We can say that AB electrodynamics is ``hidden" by strict local charge conservation, or can go unnoticed for ``small" violations thereof.

To see that ideal state-reduction events may still yield statistically conserved currents let us study the state reduction associated with the measurement of a given magnitude with corresponding operator $\widehat{A}$. We can thus expand the pre-reduction wavefunction in terms of the eigenfunctions of $\widehat{A}$:
\begin{equation}
\psi _{-}=\sum_{n}a_{n}\psi _{n},  \label{psim}
\end{equation}
where all wave functions are assumed to be normalized. In this way, each measurement yields a wave function $\psi _{+}=\psi _{n}$ with probability $\left\vert a_{n}\right\vert ^{2}$. The statistical average of equation (\ref{DmuJmu}) thus gives 
\begin{eqnarray*}
\left\langle \partial _{\mu }J^{\mu }\right\rangle &=&q\left( \left\langle
\left\vert \psi _{+}\right\vert ^{2}\right\rangle -\left\vert \psi
_{-}\right\vert ^{2}\right) \delta \left( t-t_{0}\right) \\
&=&q\left( \sum_{n}\left\vert a_{n}\right\vert ^{2}\left\vert \psi
_{n}\right\vert ^{2}-\left\vert \psi _{-}\right\vert ^{2}\right) \delta
\left( t-t_{0}\right) ,
\end{eqnarray*}
which, since 
\begin{equation*}
\left\vert \psi _{-}\right\vert ^{2}=\sum_{n}\left\vert a_{n}\right\vert
^{2}\left\vert \psi _{n}\right\vert ^{2},
\end{equation*}
results in $\left\langle \partial _{\mu }J^{\mu }\right\rangle =0$.

We can alternatively think of the above process as occurring in a system which, during a sufficiently long interval $\tau $, generates successive
similar states (\ref{psim}) so that the time average of (\ref{DmuJmu}) is ($Q$ is the amount of charge transported during $\tau$)
\begin{eqnarray*}
\overline{\partial _{\mu }J^{\mu }} &=&\frac{1}{\tau }\int_{t}^{t+\tau
}\partial _{\mu }J^{\mu }dt \\
&=&\frac{Q}{\tau }\left( \sum_{n}\left\vert a_{n}\right\vert ^{2}\left\vert
\psi _{n}\right\vert ^{2}-\left\vert \psi _{-}\right\vert ^{2}\right)
\end{eqnarray*}
for a $\tau $ large enough for all possible outcomes to be verified, resulting
again in $\overline{\partial _{\mu }J^{\mu }}=0$.

To analyze a situation in which collapse processes may lead to a persistent non-conserved current, consider a biased tunnelling structure consisting of two leads, denoted by $L$ and $R$, separated by a barrier. We focus on the barrier region and model the corresponding quantum state by the effective two-state ansatz
\begin{equation}
\psi(\mathbf{x},t)=a(t)\psi_L(\mathbf{x})+b(t)\psi_R(\mathbf{x}), \qquad |a|^2+|b|^2=1,
\label{psiprecol}
\end{equation}
where \(\psi_L\) and \(\psi_R\) are normalized wavefunctions localized, respectively, on the left and right sides of the barrier. This expression is not intended as a complete microscopic description of the open transport problem, but as a local effective ansatz for the part of the wavefunction relevant to the tunnelling event. In particular, the coefficients $a$ and $b$ encode the asymmetry produced by the external bias and by the barrier transparency. When the bias favors transport from left to right and the particle energy lies below the barrier height, one expects
\[
|a|^2>|b|^2.
\]
In the absence of collapse, the state evolves according to the appropriate Schrödinger dynamics for the effective tunnelling problem, and the corresponding Noether current is locally conserved. The point of the present argument is that, if individual tunnelling events are associated with collapse-like state reductions, then local conservation may fail at the level of single events even though a stationary average current is present.
We therefore distinguish between two different ingredients:
(i) the coherent pre-collapse state \eqref{psiprecol}, which represents the state in the barrier region immediately before an individual tunnelling event is resolved; and
(ii) the post-collapse localized outcomes, namely \(\psi_R\) for a direct transfer event and \(\psi_L\) for an inverse transfer event.
Let \(\Gamma_{L\to R}\) and \(\Gamma_{R\to L}\) denote the corresponding average rates of direct and inverse collapse events. The average direct and inverse electrical currents are then
\[
i_{L\to R}=q\Gamma_{L\to R}, \qquad i_{R\to L}=q\Gamma_{R\to L},
\]
and the net tunnelling current is
\[
i_T=i_{L\to R}-i_{R\to L}
= q\left(\Gamma_{L\to R}-\Gamma_{R\to L}\right).
\]
It is also convenient to introduce the inverse-to-direct ratio
\[
f=\frac{i_{R\to L}}{i_{L\to R}}
=\frac{\Gamma_{R\to L}}{\Gamma_{L\to R}},
\]
which, in a stationary transport regime, is expected to be controlled by the occupation imbalance of the two leads, and may be regarded phenomenologically as being of the order of the ratio of the corresponding Fermi populations at the relevant energy \cite{datta1995}.
Using equation \eqref{DmuJmu}, each direct collapse event contributes to the extra-current an amount
\[
q\left(|\psi_R|^2-|\psi|^2\right)\delta(t-t_0),
\]
while each inverse event contributes
\[
q\left(|\psi_L|^2-|\psi|^2\right)\delta(t-t_0).
\]
Averaging over many events in the stationary regime therefore gives
\[
\left\langle \partial_\mu J^\mu \right\rangle
=q\Gamma_{L\to R}\left(|\psi_R|^2-|\psi|^2\right)
+
q\Gamma_{R\to L}\left(|\psi_L|^2-|\psi|^2\right),
\]
or, equivalently,
\[
\left\langle \partial_\mu J^\mu \right\rangle
=i_{L\to R}\left(|\psi_R|^2-|\psi|^2\right)
+
i_{R\to L}\left(|\psi_L|^2-|\psi|^2\right).
\]
In terms of the net current \(i_T\) and of the ratio $f$, this may be written as
\begin{equation}
\left\langle \partial_\mu J^\mu \right\rangle=
\frac{i_T}{1-f}\left(|\psi_R|^2-|\psi|^2\right)
+
\frac{i_T f}{1-f}\left(|\psi_L|^2-|\psi|^2\right),
\label{x23}
\end{equation}
which makes explicit the separate roles of the direct and inverse channels.
For a sufficiently opaque barrier and strong forward bias one has \(f\ll 1\), and in the same regime one also expects \( |b|^2\ll |a|^2\simeq 1\). Equation \eqref{x23} then reduces to
\[
\left\langle \partial_\mu J^\mu \right\rangle
\simeq
i_T\left(|\psi_R|^2-|\psi_L|^2\right).
\]
If one further idealizes the left and right localized states as sharply concentrated on the two sides of the barrier, one obtains
\[
|\psi_L|^2 \simeq \delta(\mathbf{x}-\mathbf{x}_b), \qquad
|\psi_R|^2 \simeq \delta(\mathbf{x}-\mathbf{x}_b-\mathbf{d}),
\]
and therefore
\begin{equation}
\left\langle \partial_\mu J^\mu \right\rangle
\simeq
i_T\left[
\delta(\mathbf{x}-\mathbf{x}_b-\mathbf{d})-
\delta(\mathbf{x}-\mathbf{x}_b)
\right],
\label{I_tunnel}
\end{equation}
which coincides with the phenomenological elementary-source form introduced above.
A useful feature of the present formulation is that it separates the role of standard transport theory from that of collapse-induced non-conservation. The quantities \(\Gamma_{L\to R}\), \(\Gamma_{R\to L}\), and therefore \(i_T\), may in principle be estimated from any conventional tunnelling model for a biased junction: for instance, from a transfer-Hamiltonian description of Bardeen type, from a Landauer picture in terms of transmission probabilities \cite{datta1995}, or from a non-equilibrium Green-function treatment of the open device \cite{kadanoff_baym1962,haug2008quantum}. In all such approaches the coherent tunnelling dynamics itself remains standard. The additional hypothesis explored here is that the effective localization associated with individual transfer events produces, at the event level, a discontinuity of the wavefunction of the form considered in equation \eqref{psicollapse}, and therefore a non-vanishing source term in \(\partial_\mu J^\mu\). In this sense, standard transport theory fixes the event rates, while the collapse hypothesis determines the anomalous local source associated with each event.

The result above also helps clarify the physical meaning of the present construction. A vanishing statistical or time average of \(\partial_\mu J^\mu\), as in the case of standard projective measurements discussed previously, should not be interpreted as the absence of the effect. It only means that, in the corresponding ensemble or over the corresponding observation time, opposite event-level contributions compensate each other. The  result above concerns instead the local source associated with individual state-reduction events, or with suitable non-equilibrium sequences of such events.
In this sense, relevant observables need not be average quantities only, but may also involve fluctuations, correlations, or detector responses that are sensitive to event-by-event local non-conservation.

\section{Interaction of a gauge wave with a quantum system}
\label{AB_QS}

As studied in \cite{EPJC2023}, gauge waves are field configurations possible in AB electrodynamics which do not interact with systems in which strict local conservation of charge holds. In our previous works we have studied their possible interactions with systems in which charge is not locally conserved, only at the classical level. In the following we study how a quantum system is affected by a generic gauge wave configuration. For this we consider a system consisting in a fermion of mass $m$ interacting with a time-independent potential $V$, whose Hamiltonian is
\begin{equation}
H_{0}=-\frac{\hbar ^{2}}{2m}\nabla ^{2}+V.  \label{H0}
\end{equation}

A gauge wave has scalar and vector potentials of the generic form 
\begin{eqnarray*}
\varphi  &=&-\frac{\partial \chi }{\partial t}, \\
\mathbf{A} &=&\nabla \chi ,
\end{eqnarray*}
where the function $\chi $ satisfies D'Alembert equation
\begin{equation*}
\frac{1}{c^{2}}\frac{\partial ^{2}\chi }{\partial t^{2}}-\nabla ^{2}\chi =0,
\end{equation*}
with $c$ the speed of light in vacuum.

If the fermion particle has an electric charge $q$, in the presence of the gauge wave the Hamiltonian becomes, applying minimal coupling,
\begin{equation}
H=H_{0}-q\frac{\partial \chi }{\partial t}+\frac{i\hbar q}{m}\nabla \chi
\cdot \nabla +\frac{i\hbar q}{2m}\nabla ^{2}\chi +\frac{q^{2}}{2m}\left\vert
\nabla \chi \right\vert ^{2}.  \label{gwH}
\end{equation}

We now study how the expectation values of some operators change due to the
presence of the gauge wave, as compared with their values without the wave.
For this we use the general relation for the time evolution of the
expectation value $\left\langle O\right\rangle $ of a generic operator $O$:
\begin{equation}
\frac{d\left\langle O\right\rangle }{dt}=\left\langle \frac{\partial O}{
\partial t}\right\rangle +\frac{i}{\hbar }\left\langle \left[ H,O\right]
\right\rangle .  \label{Operatorevolution}
\end{equation}

We start with $H$ itself, for which, since $\left[ H,H\right] =0$, we have
from (\ref{gwH}) and (\ref{Operatorevolution}),
\begin{equation*}
\frac{d\left\langle H\right\rangle }{dt}=\left\langle -q\frac{\partial
^{2}\chi }{\partial t^{2}}+\frac{i\hbar q}{m}\nabla \left( \frac{\partial
\chi }{\partial t}\right) \cdot \nabla +\frac{i\hbar q}{2m}\nabla ^{2}\left( 
\frac{\partial \chi }{\partial t}\right) +\frac{q^{2}}{m}\nabla \left( \frac{
\partial \chi }{\partial t}\right) \cdot \nabla \chi \right\rangle .
\end{equation*}

We note that
\begin{eqnarray*}
\left\langle \nabla ^{2}\left( \frac{\partial \chi }{\partial t}\right)
\right\rangle  &=&\int \psi ^{\ast }\psi \nabla ^{2}\left( \frac{\partial
\chi }{\partial t}\right) d^{3}x \\
&=&\int \nabla \cdot \left[ \psi ^{\ast }\psi \nabla \left( \frac{\partial
\chi }{\partial t}\right) \right] d^{3}x-\int \nabla \left( \psi ^{\ast
}\psi \right) \cdot \nabla \left( \frac{\partial \chi }{\partial t}\right)
d^{3}x \\
&=&-\int \left( \psi ^{\ast }\nabla \psi +\psi \nabla \psi ^{\ast }\right)
\cdot \nabla \left( \frac{\partial \chi }{\partial t}\right) d^{3}x,
\end{eqnarray*}
where it was used that the first integral in the second line is zero, and also that
\begin{equation*}
\left\langle \nabla \left( \frac{\partial \chi }{\partial t}\right) \cdot
\nabla \right\rangle =\int \psi ^{\ast }\nabla \psi \cdot \nabla \left( 
\frac{\partial \chi }{\partial t}\right) d^{3}x.
\end{equation*}
We have in this way
\begin{equation*}
\frac{d\left\langle H\right\rangle }{dt}=-\int \rho \frac{\partial ^{2}\chi 
}{\partial t^{2}}d^{3}x-\int \mathbf{j}\cdot \nabla \left( \frac{\partial
\chi }{\partial t}\right) d^{3}x,
\end{equation*}
where the expressions of the charge and current densities 
\begin{eqnarray}
\rho  &=&q\psi ^{\ast }\psi ,\label{rho} \\
\mathbf{j} &=&-\frac{i\hbar q}{2m}\left( \psi ^{\ast }\nabla \psi -\psi
\nabla \psi ^{\ast }\right) -\frac{q^{2}}{m}\psi ^{\ast }\psi \nabla \chi ,\label{jc}
\end{eqnarray}
were used. 

We can thus write
\begin{eqnarray*}
\frac{d\left\langle H\right\rangle }{dt} &=&-\frac{d}{dt}\int \rho \frac{
\partial \chi }{\partial t}d^{3}x+\int \frac{\partial \rho }{\partial t}
\frac{\partial \chi }{\partial t}d^{3}x \\
&&-\int \nabla \cdot \left( \mathbf{j}\frac{\partial \chi }{\partial t}
\right) d^{3}x+\int \nabla \cdot \mathbf{j}\left( \frac{\partial \chi }{
\partial t}\right) d^{3}x,
\end{eqnarray*}
that is
\begin{equation*}
\frac{d\left\langle H\right\rangle }{dt}=-\frac{d}{dt}\left\langle q\frac{
\partial \chi }{\partial t}\right\rangle +\int \left( \frac{\partial \rho }{
\partial t}+\nabla \cdot \mathbf{j}\right) \left( \frac{\partial \chi }{
\partial t}\right) d^{3}x.
\end{equation*}

Noting further that 
\begin{equation*}
H+q\frac{\partial \chi }{\partial t}=T+V,
\end{equation*}
where the kinetic energy operator is
\begin{equation*}
T=-\frac{\hbar ^{2}}{2m}\nabla ^{2}+\frac{i\hbar q}{m}\nabla \chi \cdot
\nabla +\frac{i\hbar q}{2m}\nabla ^{2}\chi +\frac{q^{2}}{2m}\left\vert
\nabla \chi \right\vert ^{2},
\end{equation*}
we finally have for the evolution of the expectation value of the mechanical energy
\begin{equation*}
\frac{d}{dt}\left\langle T+V\right\rangle =\int \left( \frac{\partial \rho }{
\partial t}+\nabla \cdot \mathbf{j}\right) \left( \frac{\partial \chi }{
\partial t}\right) d^{3}x.
\end{equation*}

For a locally conserved charge we thus recover the result from quantum mechanics that the expectation values of physical magnitudes are gauge
independent, so that the mechanical energy for a time independent potential is conserved.

On the other hand, if the charge were not conserved locally, we recover the result from \cite{minotti2021quantum} that an additional power per unit volume of value 
\begin{equation}
-\left( \frac{\partial \rho }{\partial t}+\nabla \cdot \mathbf{j}\right)
\varphi \label{GW_power}
\end{equation}
acts on the system.

We now proceed with the canonical momentum operator $\mathbf{p}=-i\hbar
\nabla $, for which, from expressions (\ref{gwH}) and (\ref{Operatorevolution}) we readily have, using the same procedures as in the previous case,
\begin{equation*}
\frac{d\left\langle \mathbf{p}\right\rangle }{dt}=\frac{i}{\hbar }
\left\langle \left[ H_{0},\mathbf{p}\right] \right\rangle +\int \rho \nabla
\left( \frac{\partial \chi }{\partial t}\right) d^{3}x+\int \mathbf{j}\cdot
\nabla \nabla \chi d^{3}x,
\end{equation*}
which can be rewritten as
\begin{equation*}
\frac{d\left\langle \mathbf{p}\right\rangle }{dt}=\frac{i}{\hbar }
\left\langle \left[ H_{0},\mathbf{p}\right] \right\rangle +\frac{d}{dt}
\left\langle q\nabla \chi \right\rangle -\int \left( \frac{\partial \rho }{
\partial t}+\nabla \cdot \mathbf{j}\right) \nabla \chi d^{3}x.
\end{equation*}

We thus obtain the result that the evolution of the expectation value of the mechanical momentum is
\begin{equation*}
\frac{d}{dt}\left\langle \mathbf{p}-q\nabla \chi \right\rangle =\frac{i}{
\hbar }\left\langle \left[ H_{0},\mathbf{p}\right] \right\rangle -\int
\left( \frac{\partial \rho }{\partial t}+\nabla \cdot \mathbf{j}\right)
\nabla \chi d^{3}x,
\end{equation*}
and thus not affected by the presence of the gauge wave if charge is locally conserved.

For the case of charge not locally conserved we also recover the result from \cite{minotti2021quantum} that on the system acts a force per unit volume of value
\begin{equation*}
-\left( \frac{\partial \rho }{\partial t}+\nabla \cdot \mathbf{j}\right) 
\mathbf{A}.
\end{equation*}

\section{Interaction of a gauge wave with a boson quantum system}
\label{GW_Boson}

We argued recently \cite{minotti_modanese_IJMPA2026} that boson systems interacting with e.m.\ fields could have to be described by a modified Schr\"{o}dinger equation, and, correspondingly, with e.m.\ equations that should include a generally non-conserved source, in addition to the usual one. Applying that theory, if we consider the same basic Hamiltonian $H_{0}$ given by (\ref{H0}), when the particle with electric charge $q$ is in the presence of the gauge wave the Hamiltonian of the modified equation becomes
\begin{equation}
H=H_{0}+H_{1}+\frac{q^{2}}{2m}\left\vert \nabla \chi \right\vert ^{2},
\label{gwHm}
\end{equation}
where
\begin{equation}
H_{1}=-q\frac{\partial \chi }{\partial t}+\frac{i\hbar q}{m}\nabla \chi
\cdot \nabla +\frac{i\hbar q}{2m}\nabla ^{2}\chi +\frac{q^{2}}{2m}\left\vert
\nabla \chi \right\vert ^{2}.  \label{gwH1}
\end{equation}
Note that $H_{0}+H_{1}$ corresponds to the Hamiltonian in the usual Schr\"{o}dinger equation with minimal coupling. 

Proceeding as before we derive now
\begin{equation*}
\frac{d\left\langle H\right\rangle }{dt}=-\int \rho \frac{\partial ^{2}\chi 
}{\partial t^{2}}d^{3}x-\int \mathbf{j}\cdot \nabla \left( \frac{\partial
\chi }{\partial t}\right) d^{3}x,
\end{equation*}
where the expressions of the charge and current densities are now given by
\begin{eqnarray*}
\rho  &=&q\psi ^{\ast }\psi , \\
\mathbf{j} &=&-\frac{i\hbar q}{2m}\left( \psi ^{\ast }\nabla \psi -\psi
\nabla \psi ^{\ast }\right) -\frac{2q^{2}}{m}\psi ^{\ast }\psi \nabla \chi 
\\
&=&\mathbf{j}_{c}-\frac{q^{2}}{m}\psi ^{\ast }\psi \nabla \chi ,
\end{eqnarray*}
where $\mathbf{j}_{c}$ corresponds to the usual expression (\ref{jc}), which satisfies local conservation of charge when no collapse processes are involved.

As before 
\begin{equation*}
H+q\frac{\partial \chi }{\partial t}=T+V,
\end{equation*}
where the kinetic energy operator is now
\begin{equation*}
T=\frac{1}{2m}\left( \mathbf{p}-q\nabla \chi \right) ^{2}=-\frac{\hbar ^{2}}{
2m}\nabla ^{2}+\frac{i\hbar q}{m}\nabla \chi \cdot \nabla +\frac{i\hbar q}{2m
}\nabla ^{2}\chi +\frac{q^{2}}{2m}\left\vert \nabla \chi \right\vert ^{2},
\end{equation*}
so that we have for the evolution of the expectation value of the mechanical energy
\begin{eqnarray}
\frac{d}{dt}\left\langle T+V\right\rangle  &=&\int \left( \frac{\partial
\rho }{\partial t}+\nabla \cdot \mathbf{j}\right) \left( \frac{\partial \chi 
}{\partial t}\right) d^{3}x  \notag \\
&=&-\frac{q^{2}}{m}\int \nabla \cdot \left( \psi ^{\ast }\psi \nabla \chi
\right) \left( \frac{\partial \chi }{\partial t}\right) d^{3}x  \notag \\
&=&\frac{q^{2}}{2m}\int \psi ^{\ast }\psi \frac{\partial }{\partial t}
\left\vert \nabla \chi \right\vert ^{2}d^{3}x,  \label{PowerGW}
\end{eqnarray}
where the expressions in the last two lines are valid when charge is locally conserved,

Proceeding now with the canonical momentum operator $\mathbf{p}=-i\hbar\nabla $, using the same procedures as in the previous case,
\begin{equation*}
\frac{d\left\langle \mathbf{p}\right\rangle }{dt}=\frac{i}{\hbar }
\left\langle \left[ H_{0},\mathbf{p}\right] \right\rangle +\int \rho \nabla
\left( \frac{\partial \chi }{\partial t}\right) d^{3}x+\int \mathbf{j}\cdot
\nabla \nabla \chi d^{3}x,
\end{equation*}
which can be rewritten as
\begin{equation*}
\frac{d\left\langle \mathbf{p}\right\rangle }{dt}=\frac{i}{\hbar }
\left\langle \left[ H_{0},\mathbf{p}\right] \right\rangle +\frac{d}{dt}
\left\langle q\nabla \chi \right\rangle -\int \left( \frac{\partial \rho }{
\partial t}+\nabla \cdot \mathbf{j}\right) \nabla \chi d^{3}x.
\end{equation*}

We thus obtain the result that the evolution of the expectation value of the mechanical momentum is
\begin{eqnarray}
\frac{d}{dt}\left\langle \mathbf{p}-q\nabla \chi \right\rangle  &=&\frac{i}{
\hbar }\left\langle \left[ H_{0},\mathbf{p}\right] \right\rangle -\int
\left( \frac{\partial \rho }{\partial t}+\nabla \cdot \mathbf{j}\right)
\nabla \chi d^{3}x \notag \\
&=&\frac{i}{\hbar }\left\langle \left[ H_{0},\mathbf{p}\right] \right\rangle
+\frac{q^{2}}{m}\int \nabla \cdot \left( \psi ^{\ast }\psi \nabla \chi
\right) \nabla \chi d^{3}x,\label{dpdtGW}
\end{eqnarray}
where in the last line we have considered that charge is locally conserved. 

In the final expressions (\ref{PowerGW}) and (\ref{dpdtGW}) the only ``anomalous" interaction with the gauge wave results from the modification of Schr\"{o}dinger equation and the additional source in the e.m.\ equations. If current were not locally conserved, the contributions derived in the previous section should be added to the present results.

\section{Interaction of a gauge wave with a superconductor}
\label{GW_SC}

We consider a superconductor which, as a charged boson system, is described by the modified Schr\"{o}dinger equation of the previous section. In the presence of a gauge wave the superconductor then exchanges with it a power given by (\ref{PowerGW})
\begin{equation*}
W=\frac{q^{2}}{2m}\int \left\vert \psi \right\vert ^{2}\frac{\partial }{
\partial t}\left\vert \nabla \chi \right\vert ^{2}d^{3}x.
\end{equation*}

In order to obtain some numerical estimations let us consider the gauge wave component emitted by an ordinary antenna in the zone where the ordinary
electric field has magnitude $E$ with frequency $\omega $. The gauge wave
can be thus represented by the plane wave in the direction of the wave vector $\mathbf{k}$ \cite{EPJC2023}
\begin{equation*}
\nabla \chi \simeq \frac{E}{\omega }\cos \left( \mathbf{k}\cdot \mathbf{x}
-\omega t\right) .
\end{equation*}

We thus have
\begin{eqnarray*}
W &=&\frac{q^{2}E^{2}}{m\omega }\int \left\vert \psi \right\vert ^{2}\cos
\left( \mathbf{k}\cdot \mathbf{x}-\omega t\right) \sin \left( \mathbf{k}
\cdot \mathbf{x}-\omega t\right) d^{3}x \\
&=&\frac{q^{2}E^{2}}{2m\omega }\int \left\vert \psi \right\vert ^{2}\sin
\left( 2\mathbf{k}\cdot \mathbf{x}-2\omega t\right) d^{3}x.
\end{eqnarray*}

The power density (per unit of volume) is
\begin{equation*}
w=\frac{q^{2}E^{2}}{2m\omega }\left\vert \psi \right\vert ^{2}\sin \left( 2
\mathbf{k}\cdot \mathbf{x}-2\omega t\right) .
\end{equation*}

Taking for instance $E=10$ V/m, $\omega =2\pi f$, with $f=1$ MHz, and $
\left\vert \psi \right\vert ^{2}\simeq 10^{24}$ m$^{-3}$, we have
\begin{equation*}
\frac{q^{2}E^{2}}{2m\omega }\left\vert \psi \right\vert ^{2}\simeq 2\times
10^{11}\text{ W m}^{-3}.
\end{equation*}

Although this huge value seems unrealistic, we note that the power density exchanged between matter and fields in Maxwell theory, $w=\mathbf{j}\cdot 
\mathbf{E}$, presents a similar magnitude. Indeed, since 
\begin{equation*}
\mathbf{j}=-\frac{i\hbar q}{2m}\left( \psi ^{\ast }\nabla \psi -\psi \nabla
\psi ^{\ast }\right) -\frac{q^{2}}{m}\left\vert \psi \right\vert ^{2}\mathbf{
A},
\end{equation*}
and the transverse component of the electromagnetic field emitted by the
antenna is $\mathbf{E}=$ $-\partial \mathbf{A}/\partial t$, the last term of
the expression of $\mathbf{j}$ gives a contribution to $w$ of the form
\begin{equation*}
\frac{q^{2}}{2m}\left\vert \psi \right\vert ^{2}\frac{\partial }{\partial t}
\left\vert \mathbf{A}\right\vert ^{2},
\end{equation*}
which, since $\left\vert \mathbf{A}\right\vert =E/\omega $, gives a
contribution to $w$ of the same magnitude as that of the gauge wave component.

In the case of the ordinary electromagnetic wave this is not a problem because the wave is shielded out by the superconductor.

We thus see that the superconductor may also shield out the gauge wave component. The proof of this can be obtained using the equations for the e.m.\ potentials valid in the theory, which include the source additional to the usually conserved current $\mathbf{j}_{c}$:

\begin{eqnarray}
\frac{1}{c^{2}}\frac{\partial ^{2}\varphi }{\partial t^{2}} &=&\nabla
^{2}\varphi +\frac{\rho }{\varepsilon _{0}},  \label{finalphi} \\
\frac{1}{c^{2}}\frac{\partial ^{2}\mathbf{A}}{\partial t^{2}} &=&\nabla ^{2}
\mathbf{A+}\mu _{0}\mathbf{j}_{c}-\frac{\mu _{0}q^{2}}{m}\left\vert \psi
\right\vert ^{2}\mathbf{A},  \label{finalA}
\end{eqnarray}

with 
\begin{eqnarray}
\rho &=&q\left\vert \psi \right\vert ^{2},  \label{rho_cons} \\
\mathbf{j}_{c} &=&\frac{iq\hbar }{2m}\left( \psi \nabla \psi ^{\ast }-\psi
^{\ast }\nabla \psi \right) -\frac{q^{2}}{m}\left\vert \psi \right\vert ^{2}
\mathbf{A}.  \label{j_cons}
\end{eqnarray}

Using the charge conservation equation
\begin{equation*}
\frac{\partial \rho }{\partial t}+\nabla \cdot \mathbf{j}_{c}=0,
\end{equation*}
we can write for the Fourier transformed in time magnitudes ($k=\omega /c$)
\begin{eqnarray*}
\nabla ^{2}\widehat{\varphi }+k^{2}\widehat{\varphi } &=&\frac{i}{
\varepsilon _{0}\omega }\nabla \cdot \widehat{\mathbf{j}}_{c}, \\
\nabla ^{2}\widehat{\mathbf{A}}+k^{2}\widehat{\mathbf{A}} &=&-\mu _{0}
\widehat{\mathbf{j}}_{c}+\frac{\mu _{0}q^{2}}{m}\widehat{\left\vert \psi
\right\vert ^{2}\mathbf{A}}.
\end{eqnarray*}

If for simplicity we take $\left\vert \psi \right\vert ^{2}=n_{s}$ as a constant (otherwise we should include the  modified Schr\"{o}dinger equation for $\psi $) we have 
\begin{eqnarray*}
\nabla ^{2}\widehat{\varphi }+k^{2}\widehat{\varphi }+\frac{iq^{2}n_{s}}{
m\varepsilon _{0}\omega }\nabla \cdot \widehat{\mathbf{A}} &=&0, \\
\nabla ^{2}\widehat{\mathbf{A}}+\left( k^{2}-\frac{2\mu _{0}q^{2}n_{s}}{m}
\right) \widehat{\mathbf{A}} &=&0.
\end{eqnarray*}

Proposing solutions of the form $\widehat{\varphi }=\widehat{\varphi }
_{0}\exp \left( \mathbf{k}^{\prime }\cdot \mathbf{x}\right) $, $\widehat{
\mathbf{A}}=\widehat{\mathbf{A}}_{0}\exp \left( \mathbf{k}^{\prime }\cdot 
\mathbf{x}\right) $, we readily obtain 
\begin{eqnarray*}
k^{\prime } &=&k\sqrt{1-\frac{2}{k^{2}\lambda _{L}^{2}}}, \\
\widehat{\varphi }_{0} &=&\frac{c^{2}}{2\omega }\mathbf{k}^{\prime }\cdot 
\widehat{\mathbf{A}}_{0},
\end{eqnarray*}
where $\lambda _{L}$ is the London penetration depth
\begin{equation*}
\lambda _{L}=\left( \frac{\mu _{0}q^{2}}{m}n_{s}\right) ^{-2}.
\end{equation*}

We thus have a non zero $\widehat{\varphi }_{0}$ associated to the longitudinal component of $\widehat{\mathbf{A}}_{0}$ of value
\begin{equation*}
\widehat{\varphi }_{0}=c\widehat{A}_{0\Vert }\frac{1}{2}\sqrt{1-\frac{2}{
k^{2}\lambda _{L}^{2}}},
\end{equation*}
for comparison, the gauge wave in vacuum satisfies the relation $\widehat{
\varphi }_{0}=c\widehat{A}_{0\Vert }$.

We thus see that if the wave has a wavelength $\lambda $ in vacuum satisfying $\lambda >\sqrt{2}\pi \lambda _{L}$ its transverse and longitudinal components have a penetration length in the medium of value
\begin{equation*}
d=\frac{\lambda }{2\pi \sqrt{\left( \frac{\lambda }{\sqrt{2}\pi \lambda _{L}}
\right) ^{2}-1}}.
\end{equation*}

In terms of the angular frequency $\omega $ of the wave the condition for shielding by the medium, $\lambda >\sqrt{2}\pi \lambda _{L}$, is expressed
as 
\begin{equation*}
\omega <\sqrt{\frac{2q^{2}n_{s}}{\varepsilon _{0}m}},
\end{equation*}
where we see that the right-hand side has the magnitude of the plasma frequency of the medium, as with the usual e.m.\ waves cutoff in conductors.

In this way, we arrive at the conclusion that the superconductor also shields gauge waves, in contrast with ``ordinary" matter with fermionic conserved currents.

\section{Experimental proposal}
\label{experiment}

With the considerations in the previous sections we can devise some concrete experimental proposals. For instance, the biased barrier considered in Section \ref{QC_ABE} can be realized with the use of inverse-biased diodes. The resulting non-conserved current can then be used for the detection of a gauge wave, following the derivations in Section \ref{AB_QS}.

\begin{figure*}[ht]
\includegraphics[width=5cm]{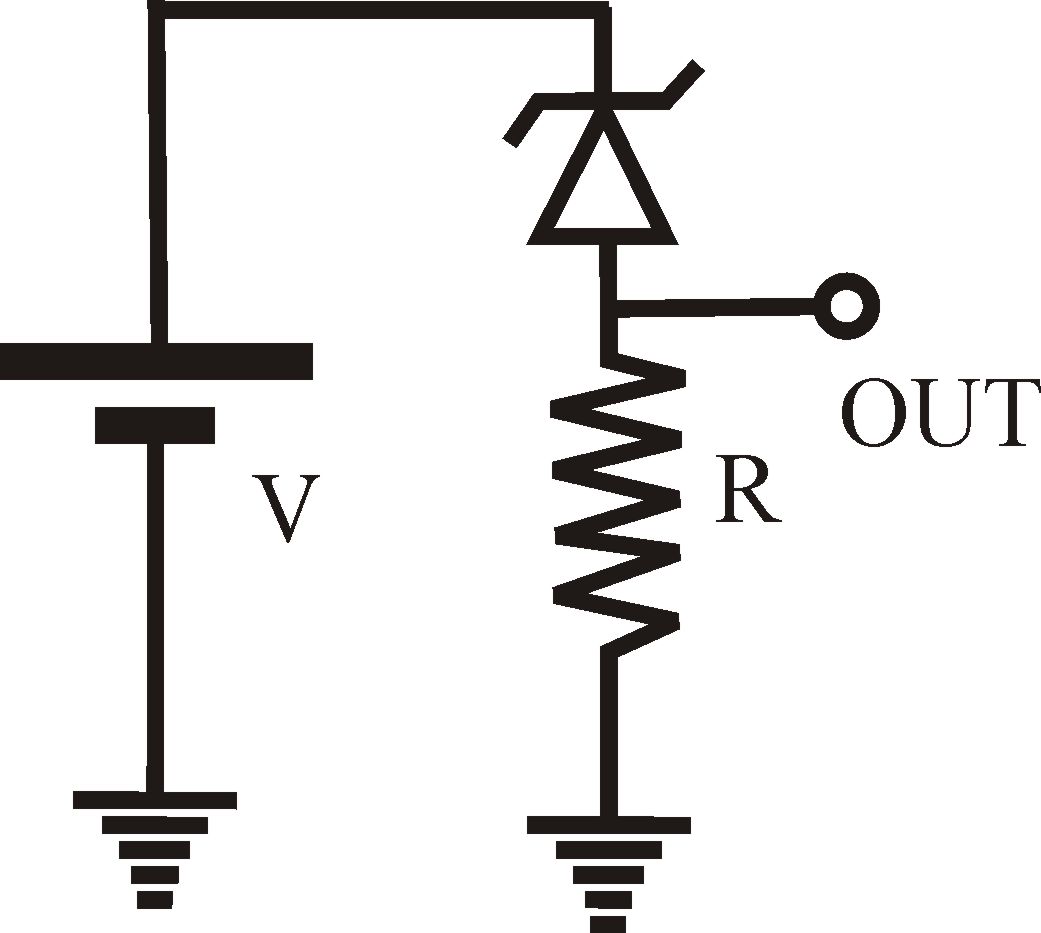}
\caption{Circuit with inverse-biased zener
diode.\label{QC1}}
\end{figure*}

In order to introduce the main ideas we can consider the simple circuit with an inverse-biased zener diode shown in fig. (\ref{QC1}).

The permanent current is determined by the relation 
\begin{equation}
i_{0}=\frac{V-\Delta V_{Z}\left( i_{0}\right) }{R},  \label{zenerI0}
\end{equation}
given the zener characteristics $\Delta V_{Z}\left( i\right) $ relating zener voltage and current. The zener characteristics, and the values of the resistance $R$ and of the DC source voltage $V$ are chosen so that the voltage across the zener $\Delta
V_{Z}\left( i_{0}\right) $ is below the zener breakdown voltage.

For a reverse-biased, low-voltage zener diode, the current $i_{0}$ is due entirely to tunneling electrons across the zener junction (high-voltage zener diodes involve additional avalanche effects). Thus, if we use expression (\ref{I_tunnel}) to quantify the violation of local conservation of charge, and expression (\ref{GW_power}) for the power per unit volume exchanged between the gauge wave and circuit, we have that the instantaneous power gained by the circuit is 

\begin{equation}
    -\int i_{0}\left[ \delta \left( \mathbf{x}-\mathbf{x}_{b}-\mathbf{d}
\right) -\delta \left( \mathbf{x}-\mathbf{x}_{b}\right) \right]\varphi\left(\mathbf{x}\right)dV=-i_{0}\Delta \varphi,
\end{equation}
where the integration is done in the volume of the junction, where tunneling occurs, and $\Delta \varphi $ is the difference of potential of the gauge wave across the junction.

Neglecting any radiated power (which is a very good approximation for
frequencies whose corresponding wavelength is large compared with the circuit dimensions), conservation of energy requires that
\begin{equation*}
i_{0}\Delta \varphi =\delta \left[ \left( \Delta V_{Z}+Ri-V\right) i\right] ,
\end{equation*}
where $\delta $ represents the variation of the magnitudes due to the interaction with the gauge wave.

For small variations about the permanent regime the small-signal analysis \cite{neamen2009} thus gives 
\begin{equation*}
i_{0}\Delta \varphi =\left[ \Delta V_{Z}\left( i_{0}\right) +Ri_{0}-V\right]
\delta i+i_{0}\left( \left. \frac{\partial V_{Z}}{\partial i}\right\vert
_{i=i_{0}}+R\right) \delta i.
\end{equation*}

Using relation (\ref{zenerI0}) we have 
\begin{equation}
\delta V_{OUT}=R\delta i=\frac{R}{\left. \frac{\partial V_{Z}}{\partial i}
\right\vert _{i=i_{0}}+R}\Delta \varphi .  \label{dVout_QC1}
\end{equation}

In this way, the circuit works as a transducer of the difference of the potential of the gauge wave across the zener junction.

\begin{figure*}[ht]
\includegraphics[width=10cm]{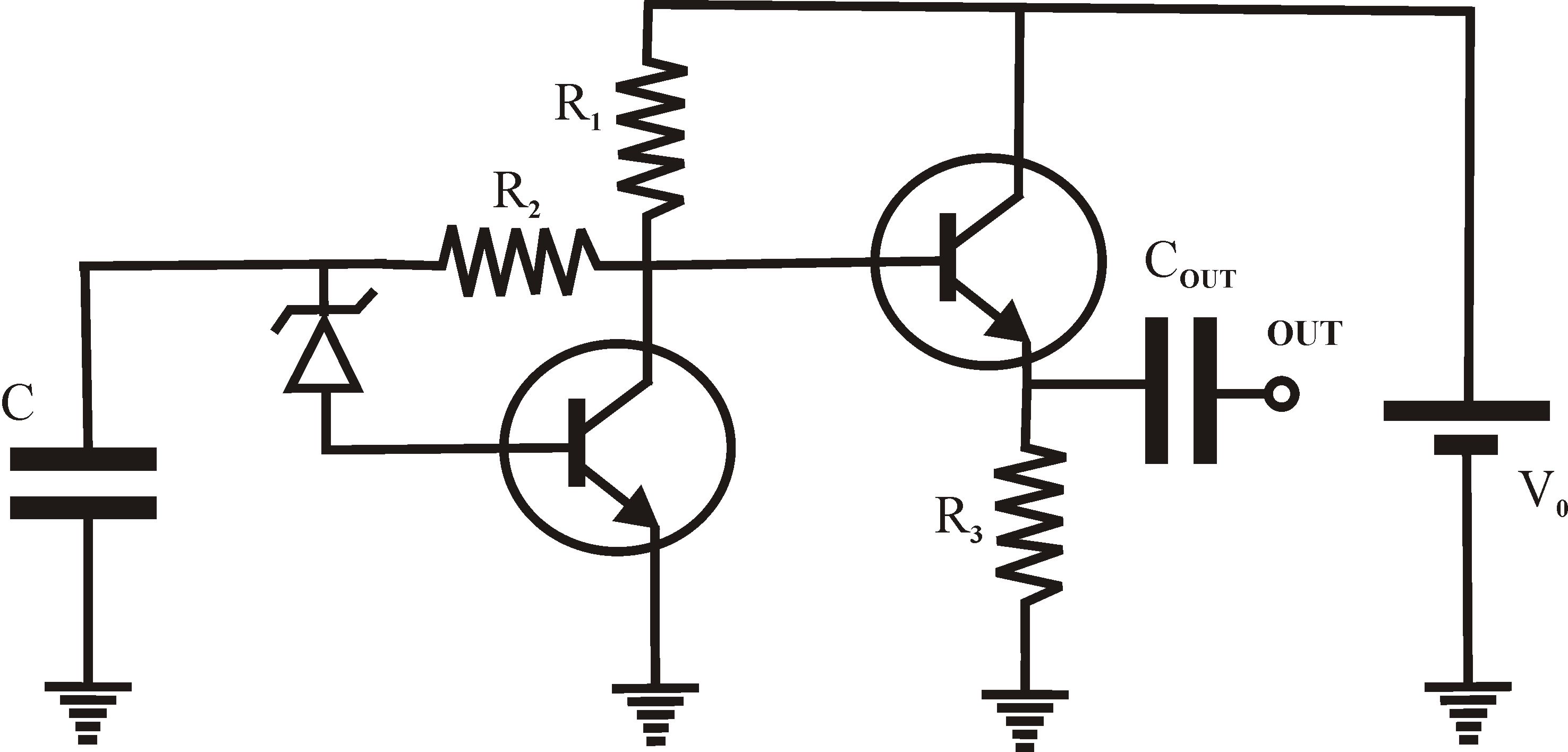}
\caption{Circuit with amplified signal of
the inverse-biased zener diode.\label{QC2}}
\end{figure*}

A more sensitive detector is given by the circuit in Fig. (\ref{QC2}). This circuit has two 2N3904 transistors \cite{2N3904} and a 1N4728A (3.3 V) zener diode \cite{1N4728A}. The other components are:
R$_{1}=\ 2.2$ k$\Omega $, R$_{2}=$ R$_{3}=10$ k$\Omega $, C $=47$ $\mu $F, C$
_{\text{OUT}}=47$ nF, V$_{0}=9$ V.

The first transistor acts as a common emitter amplifier of its base voltage, set by the zener current fluctuations. The 3.3 V zener is reverse-biased through R$_{1}$ and R$_{2}$ to about 1.64 V relative to ground, with Base-Emitter voltage of about 0.82 V (meaning a zener reverse bias of 0.82
V, which is below its breakdown voltage). Capacitor C keeps a stable voltage reference for the zener, while the second transistor acts essentially as a
voltage follower (current amplifier) that isolates the previous circuit from any external load after C$_{\text{OUT}}$.

In these conditions the current across the zener, $i_{Z},$ is about 18 $\mu $A and due entirely to tunneling electrons. In this way if a gauge wave
generates a voltage difference across the zener junction of value $\Delta\varphi $, conservation of energy, neglecting again radiation effects, can be
written as
\begin{equation}
\sum_{k}\delta \left( i_{k}\Delta V_{k}\right) =i_{Z}\Delta \varphi ,
\label{izDphi}
\end{equation}
where the index $k$ denotes the different components of the circuit, $i_{k}$ is the current through the $k$ component, $\Delta V_{k}$ the voltage across
it, and the symbol $\delta $ indicates the small variation generated by the gauge wave.

The small-signal analysis for the variations of currents and voltages about
their values at the working regime of the circuit, together with relation (\ref{izDphi}), thus determines the expected variations of all magnitudes for a given $\Delta \varphi $.

In this way, the expected voltage measured at point OUT, of value $\delta V_{OUT}=R_{3}\delta i_{3}$, with $\delta i_{3}$ the variation of the
current circulating across the resistance $R_{3}$, is determined to be
\begin{equation*}
\delta V_{OUT}\simeq 6.2\Delta \varphi .
\end{equation*}

For comparison, for the simpler circuit in Fig. (\ref{QC1}), using the same 1N4728A zener, with V $=1.5$ V, R $=10$ k$\Omega $, we have i$_{0}=3.5$ $\mu $A, which corresponds to $\left. \frac{
\partial V_{Z}}{\partial i}\right\vert _{i=i_{0}}\simeq 420$ k$\Omega $. We
thus have, according to relation (\ref{dVout_QC1}), $\delta V_{OUT}\simeq 2.2\times 10^{-2}\Delta \varphi$.

As we have shown in \cite{minotti2023aharonov,EPJC2023}, the usual electromagnetic field has a gauge component which cannot be shielded by media in which local conservation of charge holds. In this way, the circuits proposed, if effectively shielded from ordinary electromagnetic fields, can in principle detect the gauge components. 

As determined in \cite{minotti2023aharonov} all e.m.\ radiation contains in general a gauge wave component. Through interaction of the e.m.\ radiation with material media, either conductor or dielectric, the transverse Maxwell and gauge components get separated, so that a ``fog" of naturally and artificially generated gauge waves is expected to be present everywhere. Circuits like those discussed in this section would be sensitive to this ``noise", which is added to the usual thermal noise of the circuit. In this way, if two or more similar circuits, each one well shielded from ordinary e.m.\ radiation, were put in close proximity, part of the measured noise should be correlated in all circuits. By looking for correlations among the measured noise spectra in near circuits one could thus test the presented ideas.

 A rough order-of-magnitude estimate of the quantity \(\Delta\varphi\) can be obtained from the gauge-wave model used in Section V. There the gauge component is represented, in the observation region, by a potential field such that \(\nabla\chi \sim E/\omega\), where \(E\) is the amplitude of the accompanying ordinary electromagnetic field and \(\omega\) its angular frequency. For a detector dimension \(\ell\) much smaller than the wavelength, this suggests a gauge-wave potential difference of order
 $ \Delta\varphi \sim E\ell$. In the present case, however, the relevant length is not a macroscopic circuit segment, but the effective width \(d\) of the tunnelling junction entering the source term of Eq. \eqref{I_tunnel}. Taking as a reference the estimate used in Section V, namely \(E \sim 10\ {\rm V/m}\) at \(f=1\ {\rm MHz}\), one finds
 \[
 \Delta\varphi \sim E d \sim 10^{-7}\text{--}10^{-5}\ {\rm V}
 \]
 for junction widths in the approximate range \(d\sim 10\ {\rm nm}\) to \(1\ \mu{\rm m}\). This estimate is only indicative, since the actual value depends on the local gauge-wave amplitude and on the effective extent of the tunnelling region.

 It is worth comparing this with the estimate implicit in a previous detector proposal \cite{minotti2024simple}, where the relevant potential difference was taken across an entire resistive segment of the circuit, rather than across a microscopic tunnelling junction. In that case the same order-of-magnitude relation gives \(\Delta\varphi \sim E L\), with \(L\) a macroscopic length, so that \(\Delta\varphi\) may easily reach the \(10^{-2}\) to \(10^{-1}\ {\rm V}\) range for \(L\) in the millimeter-to-centimeter interval. The much smaller value expected in the present proposal is therefore not due to a weaker gauge wave, but simply to the fact that the sampled spatial scale is the microscopic junction width rather than a macroscopic circuit length.   Note, however, that although the value of $\Delta \varphi$ is in principle much larger in the case of the resistive material, the actually detected potential difference in the circuit in \cite{minotti2024simple} is reduced by the factor $\gamma$ of the material medium, expected to be of magnitude much smaller than one.

\section{Conclusions}
\label{conclusions}

In this work we have developed the idea that the usual interpretation of a quantum measurement as the collapse of the wave function, leads to an instantaneous locally non-conserved current, according to Noether theorem itself. We discuss also how the wave function collapse can lead to currents that are locally conserved or not in a statistically or time averaged sense as well.

If one accepts the presence of locally non-conserved currents, it follows that the corresponding electromagnetic fields cannot be described with Maxwell electrodynamics, which is incompatible with this non-conservation. The extended electrodynamics of Aharonov-Bohm, on the other hand, is compatible with local non-conservation of charge, and reduces to Maxwell electrodynamics when charge is strictly conserved.

Among other features, Aharonov-Bohm electrodynamics predicts the existence of particular field configurations, denoted generically as gauge waves, which only interact with non-conserved currents. In general, the e.m.\ fields generated by usual, conserved sources contain a gauge component, which goes unnoticed as long as only conserved currents are present. With this in mind, we have also studied how quantum systems interact with a gauge wave, considering the distinction between fermion and boson systems, for which a previously proposed common principle for their interaction with the e.m.\ field yields crucial differences. The principle is equivalent to minimal coupling for fermions, while for bosons the resulting equations are different.

We find that fermions interact with a gauge wave only when collapse processes lead to local non-conservation of charge, while bosons interact with that wave, even if charge is strictly conserved.

As an example of collective boson systems we consider a superconductor and show that in a similar manner as it shields ordinary e.m.\ radiation, the superconductor also effectively shields a gauge wave.

We finally discuss how a fermion system with non-conserved currents can be experimentally studied using inverse-biased diodes as systems in which electrons can give rise to non-conserved currents as they tunnel across the diode junction. Simple circuits are presented which according to the theory would be sensitive to a gauge wave, and their response to that wave quantitatively evaluated for real electronic components.

A limitation of the present treatment is that the reduction of the wavefunction has been modeled as an instantaneous state transition, represented by the step-like form of equation \eqref{psicollapse}. In this effective description, the corresponding violation of local charge conservation appears as a distributional source term proportional to $\delta(t-t_0)$, see equation \eqref{DmuJmu}. This provides a simple and transparent formulation of the basic idea, but should be regarded as an idealized kinematical model rather than as a complete dynamical description of state reduction.

It would therefore be important, in future work, to examine whether similar local non-conservation effects survive in more realistic finite-time, continuous, stochastic, or decoherence-induced reduction schemes.  In principle, for any description of the state-reduction dynamics that modifies the equations of motion derived from the original action, local non-conservation is expected, according to relation (\ref{dSzero}), with particular characteristics in each case, that require a careful study.   At the same time, the fact that the statistical average of $\partial_\mu J^\mu$ may vanish in standard projective-measurement ensembles does not make the effect physically irrelevant. It only indicates that opposite event-level contributions may cancel at the level of simple averages. From the present perspective, the most relevant observables may therefore include not only average currents, but also fluctuation-based, correlation-based, or event-induced responses of detectors sensitive to locally non-conserved sources.

More generally, the framework proposed here separates two conceptually distinct ingredients: on the one hand, the standard coherent dynamics of tunnelling or transport, which may still be described by conventional microscopic models; on the other hand, the additional hypothesis that individual transfer or measurement events involve state-reduction processes that generate a local non-vanishing source term in $\partial_\mu J^\mu$. Embedding this event-based picture into a more complete microscopic open-system description appears to be a natural continuation of the present work.

\bibliography{qc} 

@Article{minotti_modanese_IJMPA2026,
AUTHOR = {Minotti, F. and Modanese, G.},
TITLE = {Do we need an alternative to local gauge coupling to electromagnetic fields?},
JOURNAL = {International Journal of Modern Physics A},
VOLUME = {41},
YEAR = {2026},
NUMBER = {1},
PAGES = {2650024},
ARTICLE-NUMBER = {},
URL = {},
ISSN = {},
DOI = {10.1142/S0217751X26500247}
}

@Manual{2N3904,
  title        = {{2N3904/MMBT3904/PZT3904}, {NPN} {G}eneral {P}urpose {A}mplifier},
  year         = {2001},
  number       = {2N3904/MMBT3904/PZT3904},
  note         = {Rev. A},
  organization = {Fairchild Semiconductor Corporation},
  url          = {https://datasheet.octopart.com/2N3904TF-Fairchild-datasheet-8321252.pdf},
}

@Manual{1N4728A,
  title        = {{1N4728A to 1N4761A}, {Z}ener {D}iodes},
  year         = {2024},
  number       = {85816},
  note         = {Rev. 2.7},
  organization = {{V}ishay {S}emiconductors},
  url          = {https://www.vishay.com/docs/85816/1n4728a.pdf},
}

@Article{minotti_modanese_math13050892,
AUTHOR = {Minotti, F. and Modanese, G.},
TITLE = {Generalized Local Charge Conservation in Many-Body Quantum Mechanics},
JOURNAL = {Mathematics},
VOLUME = {13},
YEAR = {2025},
NUMBER = {5},
ARTICLE-NUMBER = {},
PAGES = {892},
URL = {https://www.mdpi.com/2227-7390/13/5/892},
ISSN = {2227-7390},
DOI = {10.3390/math13050892}
}

@book{haug2008quantum,
  title={Quantum kinetics in transport and optics of semiconductors},
  author={Haug, Hartmut and Jauho, Antti-Pekka and others},
  volume={2},
  year={2008},
  publisher={Springer}
}

@book{neamen2009,
  title={{M}icroelectronics {C}ircuit {A}nalysis and {D}esign},
  author={Neamen, Donald A.},
  year={2009},
  publisher={McGraw Hill}
}

@book{datta1995,
  title={Electronic transport in mesoscopic systems},
  author={Datta, Supriyo},
  year={1995},
  publisher={Cambridge University Press}
}

@book{kadanoff_baym1962,
  title={{Q}uantum {S}tatistical {M}echanics},
  author={Kadanoff, Leo P. and Baym, Gordon},
  year={1962},
  publisher={W. A. Benjamin, Inc.}
}

@book{ryder1994,
  title={{Q}uantum {F}ield {T}heory},
  author={Ryder, Lewis H.},
  year={1994},
  publisher={Cambridge, University Press}
}

@article{padmanabhan2018,
  title={Obtaining the non-relativistic quantum mechanics from quantum field theory: issues, folklores and facts.},
  author={Padmanabhan, T.},
  journal={The European Physical Journal C},
  volume={78},
  pages={563},
  year={2018},
  publisher={Springer Nature}
}

@book{huang2010quantum,
  title={Quantum field theory: From operators to path integrals},
  author={Huang, Kerson},
  year={2010},
  publisher={John Wiley \& Sons}
}

@article{EPJC2023,
  title={Gauge waves generation and detection in {A}haronov–{B}ohm electrodynamics},
  author={Minotti, Fernando and Modanese, Giovanni},
  journal={The European Physical Journal C},
  volume={83},
  number={ },
  pages={1086},
  year={2023},
  publisher={Springer Nature}
}

@article{minotti2021quantum,
  title={Quantum uncertainty and energy flux in extended electrodynamics},
  author={Minotti, Fernando and Modanese, Giovanni},
  journal={Quantum Reports},
  volume={3},
  number={4},
  pages={703--723},
  year={2021},
  publisher={Multidisciplinary Digital Publishing Institute}
}

@article{minotti2024simple,
  title={Simple circuit and experimental proposal for the detection of gauge-waves},
  author={Minotti, Fernando and Modanese, Giovanni},
  journal={Journal of Physics Communications},
  volume={8},
  number={5},
  pages={055003},
  year={2024},
  publisher={IOP Publishing}
}

@article{minotti2023aharonov,
  title={{Aharonov--Bohm Electrodynamics in Material Media: A Scalar e.m. Field Cannot Cause Dissipation in a Medium}},
  author={Minotti, Fernando and Modanese, Giovanni},
  journal={Symmetry},
  volume={15},
  number={5},
  pages={1119},
  year={2023},
  publisher={MDPI}
}

@article{minotti2022electromagnetic,
  title={Electromagnetic Signatures of Possible Charge Anomalies in Tunneling},
  author={Minotti, Fernando and Modanese, Giovanni},
  journal={Quantum Reports},
  volume={4},
  number={3},
  pages={277--295},
  year={2022},
  publisher={MDPI}
}

@article{hively2019classical,
  title={Classical and extended electrodynamics},
  author={Hively, L.M. and Loebl, A.S.},
  journal={Physics Essays},
  volume={32},
  number={1},
  pages={112-126},
  year={2019},
  publisher={Physics Essays Publications}
}

@article{aharonov1963further,
  title={Further discussion of the role of electromagnetic potentials in the quantum theory},
  author={Aharonov, Y. and Bohm, D.},
  journal={Physical Review},
  volume={130},
  number={4},
  pages={1625},
  year={1963},
  publisher={APS}
}

@article{modanese2018time,
  title={Time in quantum mechanics and the local non-conservation of the probability current},
  author={Modanese, G.},
  journal={Mathematics},
  volume={6},
  number={9},
  pages={155},
  year={2018},
  publisher={Multidisciplinary Digital Publishing Institute}
}

@article{hively2012toward,
  title={Toward a more complete electrodynamic theory},
  author={Hively, L.M. and Giakos, G.C.},
  journal={International Journal of Signal and Imaging Systems Engineering},
  volume={5},
  number={1},
  pages={3--10},
  year={2012},
  publisher={Inderscience Publishers}
}

@ARTICLE{Modanese2017MPLB,
author={Modanese, G.},
title={{Generalized Maxwell equations and charge conservation censorship}},
journal={Modern Physics Letters B},
year={2017},
volume={31},
pages={1750052},
doi={},
url={},
document_type={Article},
source={},
}

@ARTICLE{Minotti-Modanese-Symmetry2021,
author={Minotti, F. and Modanese, G.},
title={{Are Current Discontinuities in Molecular Devices
Experimentally Observable?}},
journal={Symmetry},
year={2021},
volume={13},
pages={691},
doi={},
url={},
document_type={Article},
source={},
}
\bibliographystyle{ieeetr}

\end{document}